\documentstyle[11pt,paspconf,psfig, times]{article}

\include{psfig.sty}

\begin{document}

\title{Carbon stars in the Local Group}
\author{M.A.T. Groenewegen}
\affil{Instituut voor Sterrenkunde, KU Leuven, Celestijnenlaan 200B, \\ 
B-3001 Leuven, Belgium}

\begin{abstract}
This is the unofficial proceeding of my invited review at the Ringberg
Castle Workshop on ``The Chemical Evolution of Dwarf Galaxies'', July
28th - August 2nd, 2002, and presents an update of a similar review
I gave in 1999.  \\

\noindent
The current status of carbon stars in the Local Group and beyond, is
discussed. Although many carbon stars and late M-stars have been
identified in Local Group galaxies, a coherent understanding in terms
of the chemical evolution- and star formation history of a galaxy is
still largely lacking. Although a few major new surveys have been
carried out over the last three years, the observational data is still
incomplete in many respects: 1) for some of the larger galaxies only a
small fraction in area has been surveyed so far, 2) surveys have been
conducted using different techniques, and some of the older surveys
are incomplete in bolometric magnitude, 3) only for some galaxies is
there information about the late M-star population, or it is sometimes 
unpublished even when the data is available, 4) not all galaxies in
the Local Group have been surveyed, 5) especially for some of the
older work insufficient data is available to determine bolometric
magnitudes.  I have correlated carbon star positions with the 2MASS
2nd incremental data release to obtain $JHK$, and bolometric
magnitudes, to remedy this situation in some cases.

From the existing observations one can derive the following: the
formation of carbon stars is both a function of metallicity and
star-formation history. In galaxies with a similar star formation
history, there will be relatively more carbon stars formed in the
system with the lower metallicity. On the other hand, the scarcity of
AGB type carbon stars in some galaxies with the lowest metallicity
indicates that these systems have had a low, if any, star-formation
over the last few Gyrs.

\end{abstract}


%\vspace{-6mm}
\section{Introduction}

Carbon stars are tracers of the intermediate age population in
galaxies. Either they are currently undergoing third dredge-up on the
(Thermal-Pulsing) Asymptotic Giant Branch (TP-AGB) -- the cool and
luminous N-type carbon stars --, or have been enriched with
carbon-rich material in a binary system when the present-day white
dwarf was on the AGB (the carbon dwarfs and CH-stars. The R-stars may
be the result of a coalescing binary [McClure 1997]).

Since their spectral signature is very different from oxygen-rich and
S-type stars, it is relatively easy to identify carbon stars even at
large distances.  In Sect.~2 the main technique to identify carbon
stars is briefly discussed, and in Sect.~3 the various surveys for
carbon stars in external galaxies are summarised. That section is in
fact, an update of a review I gave at IAU Symposium 191 (Groenewegen
1999; hereafter G99), and here I will only refer to new results
published since then, or older literature when I use it to obtain new
results. Please refer to G99, and the similar review by Azzopardi
(1999) for the full story. The results and some recent theoretical
predictions are discussed in Sect.~4.

\section{Methods}

This has been discussed in some detail in G99. Briefly, the most
effective method for large scale surveys of late-type M- and C-stars
uses typically two broad-band filters from the set $V,R,I$, and two
narrow-band filters near 7800 and 8100 \AA, which are centred on a
CN-band in carbon stars, and a TiO band in oxygen-rich stars,
respectively. In an [78-81] versus $[V-I]$ (or similar) colour-colour
plot, carbon stars and late-type oxygen-rich stars clearly separate
redwards of $V-I \approx$ 1.6.  For an illustration of this, see Cook
\& Aaronson (1989), or Kerschbaum et al. (1999).

A caveat is that, unfortunately, not all groups adopt the same
``boxes'' in these diagrams to select M- and C-stars (e.g. compare
Nowotny et al. 2001 and Demers et al. 2002). Furthermore, in many
cases, only the photometry is published for the stars the respective
authors consider to be the M- and C-stars, so its impossible to apply
ones own selection criteria a posteriori. Furthermore, one usually
applies the same lower-limit on the broad band colour ($(V-I)$, or
$(R-I)$) to select M-stars (usually chosen to correspond to M0 and
later) to  C-stars, while it is known (e.g.  Kerschbaum et
al. 1999) that the hottest C-stars (spectral type C0) are bluer than
this limit.  So, applying the same lower-limit on the broad-band
colour will bias against the hottest C-stars.

\section{Surveys}

In this section the (recent) surveys for carbon stars in external
galaxies, or new results in general, are described.

\underline{The Magellanic Clouds} 

\noindent
A huge new survey of the LMC was presented by Kontizas et al. (2001)
covering about 300 sq.degree and containing 7760 objects. I used
the listed $R,I$ data to calculate the bolometric magnitude using
bolometric corrections (BCs) from Costa \& Frogel (1996; hereafter CF)
for a reddening of $E(B-V)$ = 0.06.

Kunkel et al. (2000) present a new survey of carbon stars in the
periphery of the SMC, and present $B$, $R$, and heliocentric velocities
for 150 stars.

It should be mentioned that the data release of the infra-red surveys
DENIS and 2MASS has allowed to identify many candidate AGB stars based
on colour criteria, e.g. Egan et al. (2001), who combine 2MASS with
MSX data, to suggest about 500 candidate AGB stars, or Nikolaev \&
Weinberg (2000) who analyse 2MASS data to find that there is a
significant population of AGB stars.  In the near future multi-object
spectroscopy will assign spectral types and measure radial velocities
for many of them.

\begin{figure}[t]
\centerline{\psfig{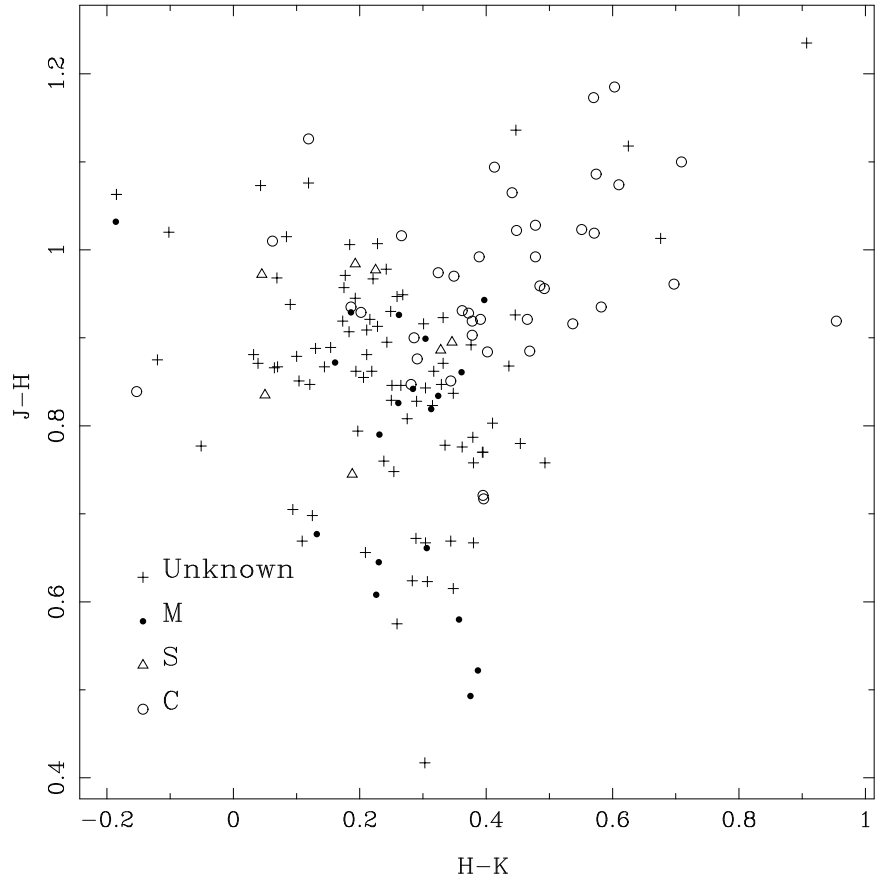}}
\caption[]{Colour-colour diagram of 152 stars in Fornax from Stetson
et al. (1998) with 2MASS counterparts, with spectral types indicated.}
%\vspace{-3mm}
\end{figure}

\begin{figure}[t]
\centerline{\psfig{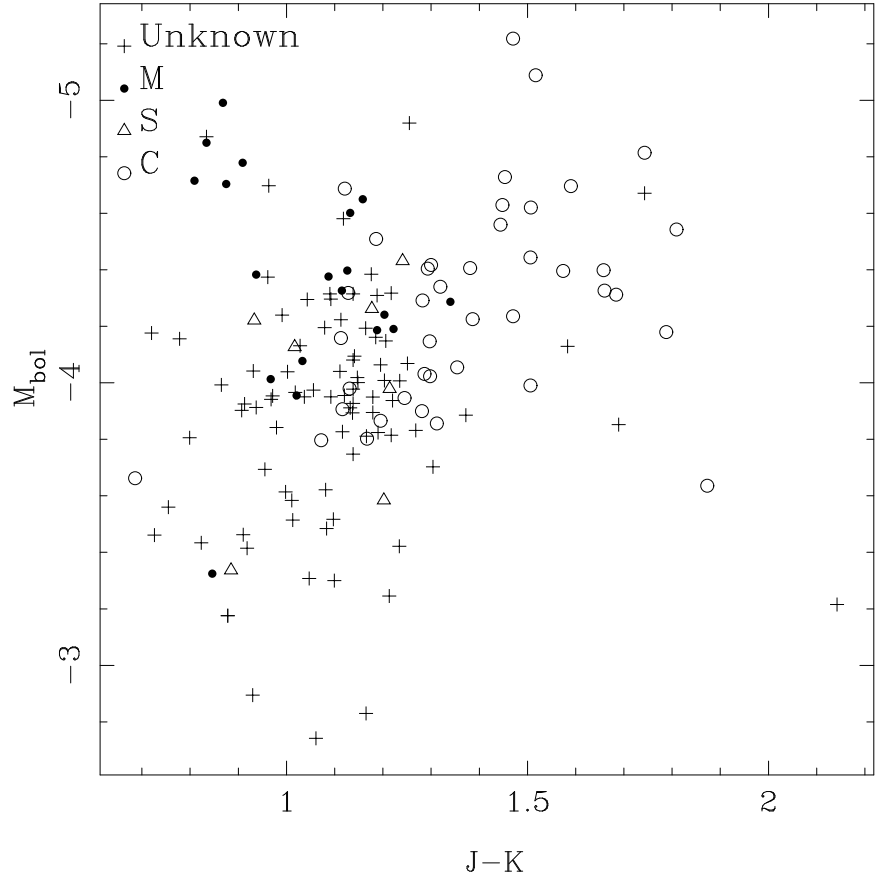}}
\caption[]{Colour-magnitude diagram for the stars from the previous
figure.}
%\vspace{-2mm}
\end{figure}

\

\newpage

\

\underline{Sagittarius dwarf galaxy}

\noindent
The total number of known carbon stars is 26, with an estimated total
of about 100 (Whitelock et al. 1999). $JHK$ photometry is listed for
11 of them in Whitelock et al. (1996). I correlated all 26 of them
with 2MASS, finding 14 matches, 9 also with photometry in Whitelock et
al. (1996). The object with the largest difference in $K$, is the
object classified by Whitelock et al. (1999) as a Mira, actually the
first confirmed Mira-type variable in a dwarf spheroidal. The
bolometric magnitudes for the 16 stars are calculated adopting
$E(B-V)$ = 0.15 and BCs from Bessell \& Wood (1984; hereafter BW84).

\underline{Fornax dwarf galaxy}

\noindent
Azzopardi et al. (1999) lists a total known number of 104, with
an estimated total of 120.

Demers \& Irwin (1987) identify 30 long-period variables but none with
Mira-type amplitudes. Bersier \& Wood (2002) give a list of 85
candidate LPVs.

Stetson et al. (1998) present a list of 161 candidate luminous red
stars in a field of 1274 sq.arcmin based on $B$ and $R$ exposures. 
Cross-identifications with Demers \& Kunkel (1979), Westerlund et
al. (1987) and Lundgren (1990) are also given.

Demers et al. (2002) consider the 2MASS observations in the direction
of Fornax and the LMC and use a selection in $J-K$ to identify 5 new
candidates, and confirm 21 known carbon stars in Fornax.

I took the coordinates from Stetson et al. and made a correlation with
the 2MASS catalogue in 2 steps. First using a search radius of 2 arcsec,
finding 140 matches. Plotting the offsets between the input
coordinates and the 2MASS coordinates revealed that there is an offset
of 0.66 arcsec in RA and 1.10 arcsec in Dec. In a second pass, the
Stetson et al. coordinates were corrected for this offset, and a search
radius of 1.5 arcsec was used. 152 matches were found, with an rms in
the difference between 2MASS and input coordinates now 0.42 arcsec in
RA and 0.32 arcsec in Dec. This sub-sample of 152 stars contains 41
C-, 7 S- and 19 M-stars.  Bolometric magnitudes are calculated
adopting a reddening $E(B-V)$ of 0.020, and BCs to the $K$ magnitude
from BW84.

Figure~1 and 2 show the colour-colour and colour-magnitude diagram for
these 152 red stars with 2MASS counterparts. From both diagrams it is
clear that the carbon stars stand out from the M- and S-stars, a
feature that is well known and related to the different molecular
absorption features in the NIR. The carbon stars are on average more
luminous than the M- and S-stars. Note that the two reddest stars seem to
have a relatively low luminosity again. This is likely to be an
artifact, since these red colours are influenced by circumstellar
reddening because of mass loss, and hence a simple BCs formulation to
obtain $M_{\rm bol}$ may become inaccurate.

%Westerlund et al. (1987) list bolometric magnitudes for 43 carbon
%stars based on $B,V$ photometry.

\begin{table}[t]
\caption[]{$(R-I)_0$ and $(V-I)_0$ colours of M-stars (with no mass
loss) derived from the spectra in Fluks et al. (1994)}

\begin{flushleft}
\begin{tabular}{lcclcc} \hline
Type & $(R-I)_0$ & $(V-I)_0$  & Type & $(R-I)_0$ & $(V-I)_0$  \\ \hline
M0 & 0.82 & 1.63 & M1 & 0.94 & 1.79 \\
M2 & 1.02 & 1.90 & M3 & 1.16 & 2.07 \\
M4 & 1.38 & 2.34 & M5 & 1.70 & 2.77 \\
M6 & 2.00 & 3.21 & M7 & 2.24 & 3.64 \\
M8 & 2.27 & 3.94 & M9 & 2.38 & 4.21 \\
M10 & 2.52 & 4.36 &  &  &  \\ 
\hline
\end{tabular}
\end{flushleft}
%\vspace{-3mm}
\end{table}

%\vspace{-3mm}
\begin{table}[t]
\caption[]{The carbon star census}
%\vspace{-2mm}
\small

\begin{flushleft}
\begin{tabular}{lrrrrlll} \hline
Name      & D    & $M_{\rm V}$ & $[$Fe/H$]$ & $N_{\rm C}$ & Area & Factor$^a$ & $N_{\rm M}$ \\ 
          & (kpc)& (mag)       &            &             & (kpc$^2$) &   &   \\ \hline
M31       &  770 & --21.2      &   0.0      & 243         & 12.3 & 47 & 789 (5+) \\
Galaxy    &      & --20.6      &   0.0      & 81          & 1.0  & 700 & C/M $\approx$ 0.2$^b$ \\
M33       &  840 & --19.0      & --0.6      & 15          & 0.20 & 790 & 5 (5+), 60 (0+) \\
LMC       &   50 & --18.1      & --0.6      & 1045        & 4.8  & 11 & 1300 (5+) \\
          &      &             &            & 7750        & 300. &   & \\
NGC 205   &  815 & --16.4      & --0.8      & 7           & 0.33 & 40 & 4 (5+), 17 (0+) \\
SMC       &   63 & --16.2      & --1.0      & 789         & 5.4  & 4 & 180 (5+)\\
          &      &             &            & 1707        & 12.2 &  & \\
NGC 6822  &  490 & --16.0      & --0.7      & 904         & 4.5  & 1.0 & 941 (0+) \\
IC 1613   &  720 & --15.3      & --1.3      & 195         & 7.8  & 1.0 & 35 (5+), 300 (0+) \\
WLM       &  930 & --14.4      & --1.5      & 14          & 0.28 & 8 & 0 (5+), 6 (0+) \\
SagDSph   &   24 & --13.4      & --1.0      & 26          & 7.2  & 4 &    \\
Fornax    &  138 & --13.1      & --1.3      & 104         & 1.35 & 1.0 & 4 (5+), 25 (2+)\\
Pegasus   &  955 & --12.4      & --1.1      & 40          & 1.04 & 1.0 & 77 (0+) \\
SagDIG    & 1060 & --12.3      & --2.5      & 16          & 0.58 & 1.0 & 1 (0+) \\
Leo I     &  250 & --11.9      & --2.0      & 23          & 0.45 & 1.0 &  1 (5+), 15 (0+) \\
And I     &  790 & --11.9      & --1.5      & 0           & &  & \\
And II    &  680 & --11.8      & --1.5      & 8           & &      & 1 (0+) \\
Aquarius  &  950 & --10.9      & --1.9      & 3           & 0.18 & 1.0 & 1 (0+) \\ 
And III   &  760 & --10.3      & --2.0      & 0           & & & \\
Leo II    &  205 & --10.1      & --1.9      & 8           & 0.47 & 1.0  & \\
Sculptor  &   79 &  --9.8      & --1.8      & 8           & 0.65 & 1.0  & 40 (2+), 0 (5+)\\
Phoenix   &  400 &  --9.8      & --1.8      & 2           & 0.40 & 1.0 & \\
Sextans   &   86 &  --9.5      & --1.7      & (0)         &      &     & \\
Carina    &  100 &  --9.4      & --2.0      & 11          & 0.31 & 1.0 & \\
Tucana    &  870 &  --9.3      & --1.8      & 0           & 0.22 & 1.0 & \\
Ursa Minor&   69 &  --8.9      & --2.2      & 7           & 0.58 & 1.0 & \\
Draco     &   82 &  --8.6      & --2.0      & 6           & 0.50 & 1.0 & \\ \hline
NGC 2403  & 3390 & --20.4      &   0.0      & 4           & 2.0  & 125 & 7 (0+) \\
NGC 300   & 2170 & --18.7      & --0.4      & 16          & 3.2  & 31  & 23 (0+), 6 (5+) \\
NGC 55    & 1480 & --18.0      & --0.6      & 14          & 2.8  & 12  & 6 (5+) \\
\hline
\end{tabular}

$^a$. Ratio of the area of the galaxy on the sky (taken from the NED catalog) 
and the survey area. \\

$^b$. In the solar neighbourhood. 

\end{flushleft}
%\vspace{-6mm}
\end{table}

\underline{Leo {\sc i} dwarf galaxy}

\noindent
The total number of known carbon stars is 23. Demers \& Battinelli
(2002) use the narrow-band filter technique to find 7 new carbon
stars. The number of M0+ stars they estimate to be 15 (foreground
corrected), and based on the expected colours of M-stars I estimated
that the number of M2+, M3+ and M5+ stars is 13, 10 and 1, respectively. 
This (and similar estimates described below for other galaxies) is
based on colours derived from the M-type spectra of Fluks et
al. (1994), as listed in Table~1.  Demers \& Battinelli (2002) give
$R,I$ for all 23 stars, and I calculated bolometric magnitudes
assuming $E(B-V)$ = 0.02 and BCs from CF.

Recently, Menzies et al. (2002) surveyed a field of 7.2 arcmin square
in the inner part of Leo {\sc i} in $JHK$. They find all 21 known and
suspected carbon stars in the field, but also identify 3 very red
objects (and a fourth in a nearby field), which, based on their
colours, are possibly carbon stars undergoing mass loss.

\underline{Sculptor dwarf galaxy}

\noindent
I took the coordinates from Azzopardi et al. (1986) and correlated
them with 2MASS to find a match in all cases. For the 2 stars with IR
photometry listed in Frogel et al. (1982) the agreement is excellent.
Bolometric magnitudes are calculated assuming $E(B-V)$ = 0.02, and BCs
from BW84. The number of stars available to calculate the LF is
therefore increased from 2 to 8.

\underline{Leo {\sc ii} dwarf galaxy}

\noindent
The total known number of certain carbon stars is 8. Azzopardi et
al. (2000) state they found 2 new ones, without providing further details.

I extracted the coordinates for the 6 certain and 1 candidate carbon
stars in Azzopardi et al. (1985), and cross-correlated them with
2MASS, to find 6 matches (all but ALW 5). The $K$-magnitudes for ALW 4
and 6 agree to with a few hundredth of a magnitude with those listed
in Aaronson \& Mould (1985), the difference in the sense Aaronson \&
Mould minus 2MASS is 0.1 mag for ALW1, --0.18 for ALW3 and --0.36 for
ALW7, possibly indicating they are variable. I determined the
bolometric magnitude for ALW2, and removed the star DH 260 in
Aaronson \& Mould (1985), which I inadvertently had included in 1999 in the
luminosity function of this galaxy, but is not a carbon star.

\begin{figure}[t]
\centerline{\psfig{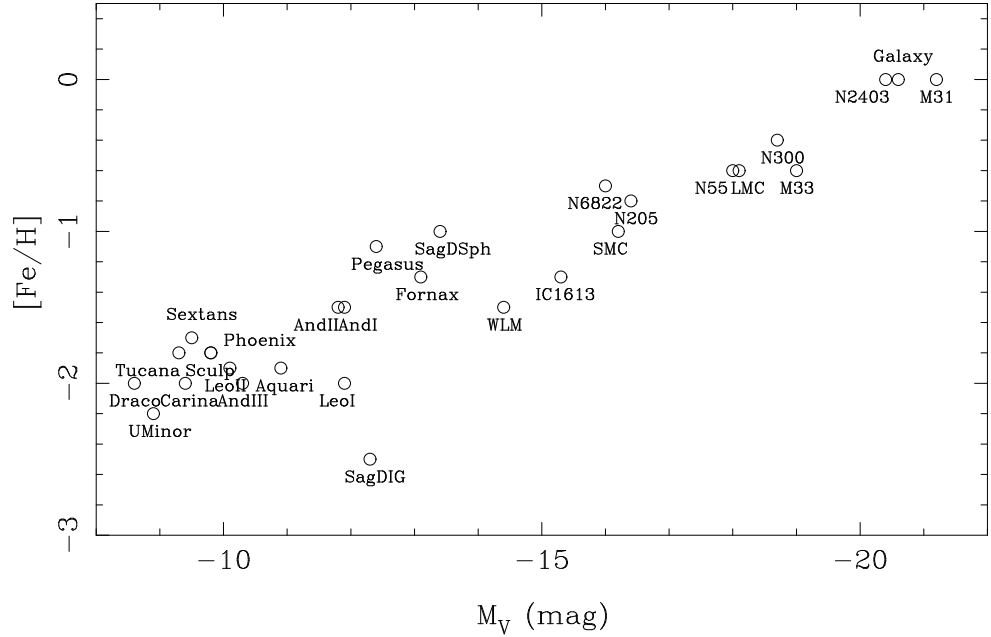}}
\caption[]{Metallicity  versus $M_{\rm V}$ of the galaxies discussed here.}
\vspace{-0mm}
\end{figure}

\begin{figure}[t]
\centerline{\psfig{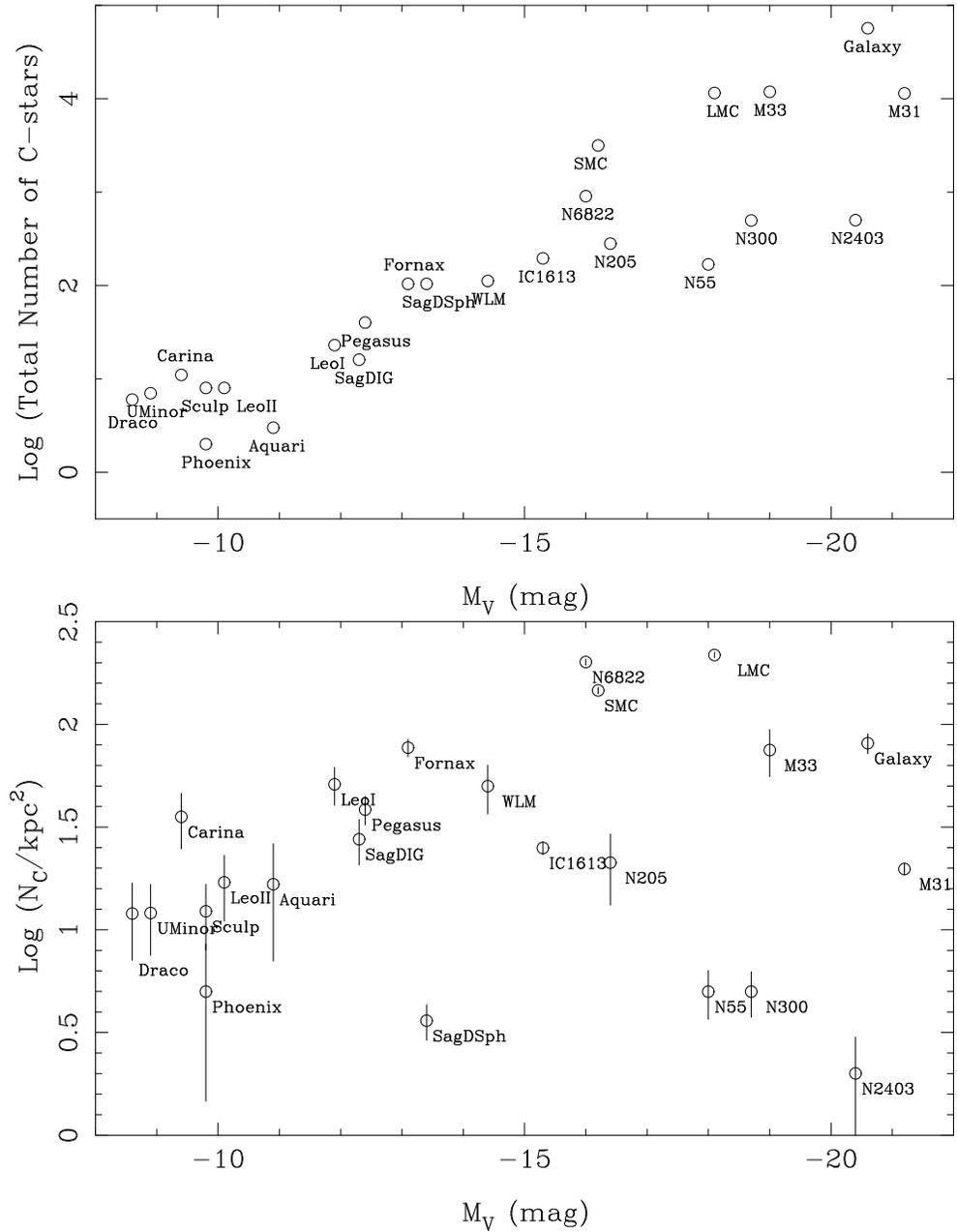}}
\caption[]{Total number of carbon stars and surface density of carbon
stars versus $M_{\rm V}$. No correction for
inclination effects has been made.}
\vspace{-4mm}
\end{figure}

\begin{figure}[t]
\centerline{\psfig{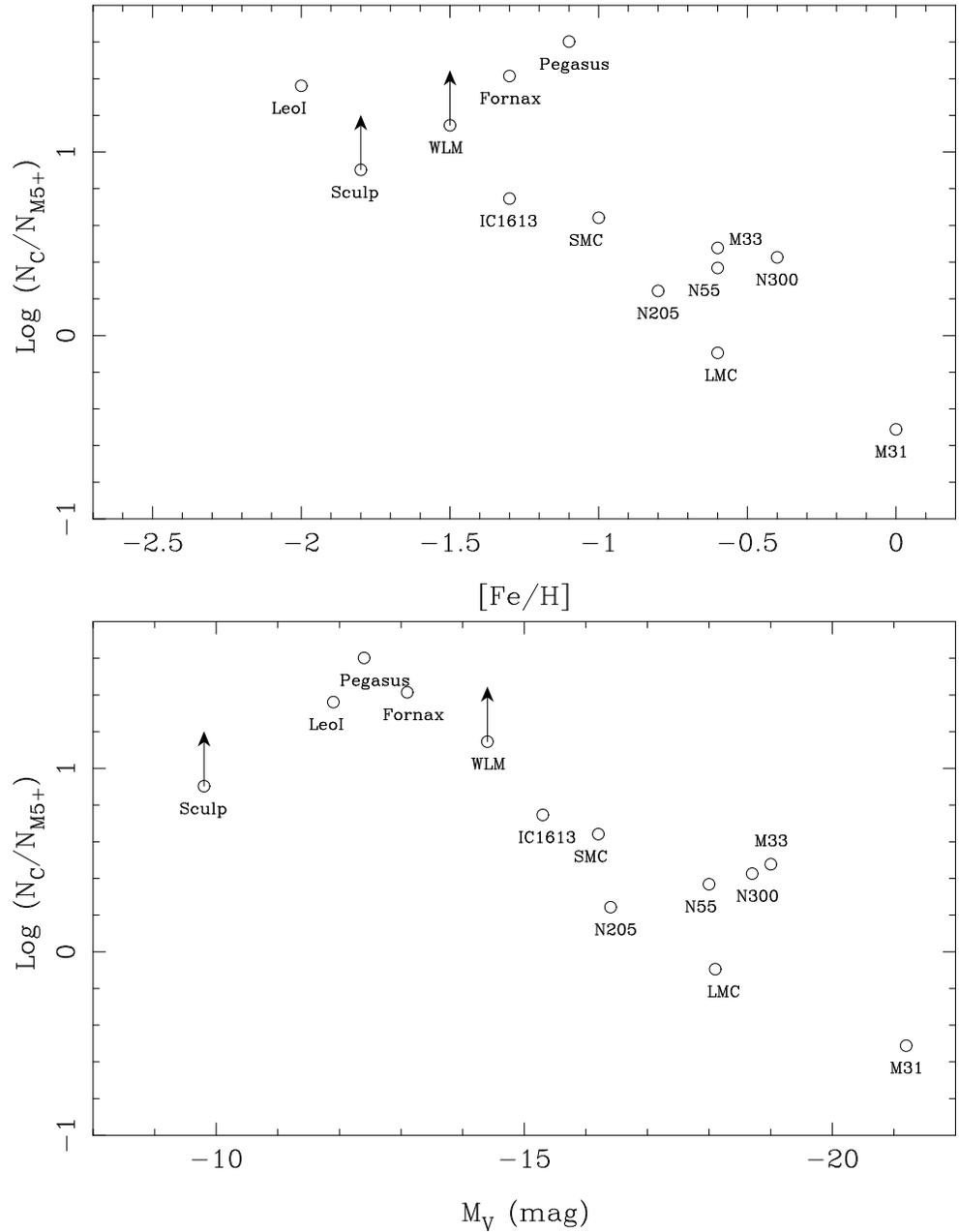}}
\caption[]{Log of the number ratio of carbon stars to late M-stars
versus metallicity and $M_{\rm V}$. }
\end{figure}

\begin{figure}[t]
\centerline{\psfig{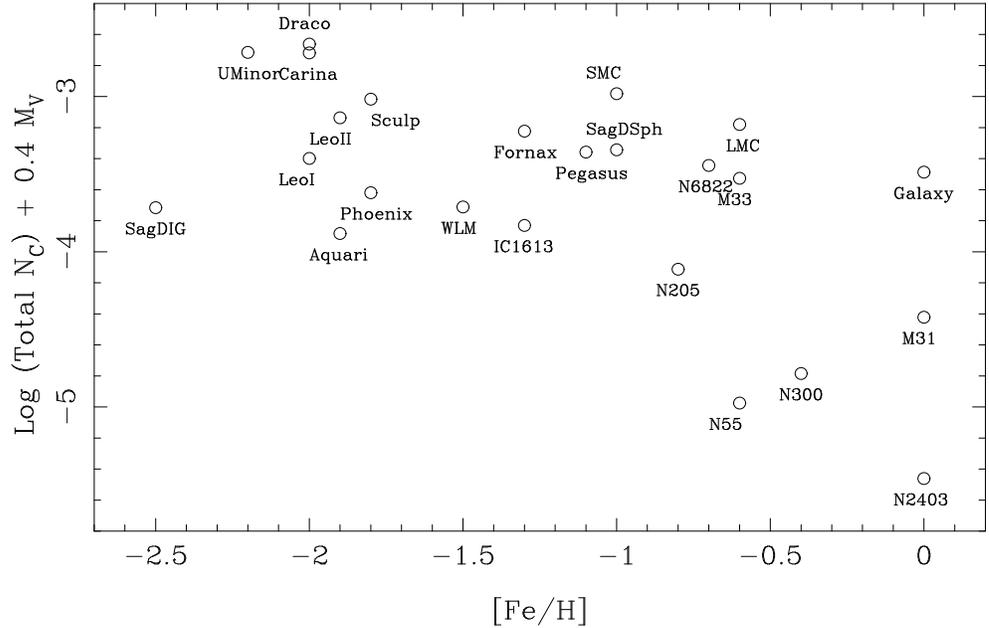}}
\caption[]{Log of the number of carbon stars over total visual
luminosity versus metallicity.}
\end{figure}

\underline{Phoenix dwarf galaxy}

\noindent
Martinez-Delgado et al. (1999) present $I$-band photometry of the two
known carbon stars that agrees very well with the values in Van de
Rydt et al. (1991).

\underline{Carina dwarf galaxy}

\noindent
Azzopardi et al. (1999) list 11 carbon stars, which is two more than
listed in G99, but no details are given.

\underline{Ursa Minor dwarf galaxy}

\noindent
The total number of known carbon stars is 7.  Armandroff et al. (1995)
confirm the 1 certain and 1 candidate carbon star in Azzopardi et
al. (1986, and references therein), and find 2 new ones.  Shetrone et
al. (2001) do follow-up spectroscopy of red stars and find 3 new
carbon stars to bring the total to 7. They list $B,V$ photometry for
all.  I correlated these 7 using the positions in Shetrone et al. with
2MASS and found a match within 1 arcsec in 6 cases, of which 5 have
good photometry in $J$ and $K$. A reddening of $E(B-V)$ = 0.032 is
adopted to compute $m_{\rm bol}$ with BCs from BW84.

\underline{Draco dwarf galaxy}

\noindent
The total known number of carbon stars is 6. Armandroff et al. (1995)
confirm the 3 certain and 1 candidate carbon star in Azzopardi et
al. (1986, and references therein), and find one new one.  Shetrone et
al. (2001) do follow-up spectroscopy of red stars and find a sixth
carbon star. $B,V$ photometry is listed for all. Draco is not yet
observed by 2MASS. Margon et al. (2002), in a search for ``Faint High
Latitude Carbon Stars'', identify three known carbon stars in SDSS data.

\underline{M31}

\noindent
From Brewer et al. (1995, table 8), and the expected colours of
M-stars (Table~1), I estimated the number of M0+, M2+, M3+, M5+, M6+
stars to be 5254, 4887, 3920, 789 and 223, respectively.

A new survey was presented by Nowotny et al. (2001) for one field of
5.3x5 arcmin (unvignetted area) in between fields 3 and 4 of Brewer et
al. (1995). They find 61 carbon stars and 507 M0+ stars. Since no
colour information for the M-stars is given, no spectral subdivision
according to colour can be made. The C/M ratio and the number of
objects per unit area found by Nowotny et al. (2001) are very similar
to those found by Brewer et al. (1995) for their flanking field no 4.

\underline{M33}

\noindent
The colours of the M-stars in Cook et al. (1986) were inspected and I
estimated the number of M0+, M2+, M3+, M5+, M6+ stars to be 60, 34,
26, 5, 2.

\

\newpage

\

\underline{And {\sc i,ii,iii}}

\noindent
Cote et al. (1999) took spectra of red stars in And {\sc ii} and
confirm one new carbon star, and give $V,I$ photometry. Da Costa et
al. (2000) quote a $V$-magnitude fainter by 0.2 mag for this object.
However, since the list of Cook et al. remains unpublished it is
unclear whether this star is in their list already. Bolometric
magnitudes are available for 2 certain C-stars, one based on IR
photometry (Aaronson et al. 1985b), and the other calculated by me
using the average of the two available $V$ data points, $V-I$ from
Cote et al. (1999), assuming $E(B-V)$ = 0.06, and BCs from BW84. \\

\

\newpage

\

\underline{NGC 6822}

\noindent
A new survey using the narrow-band filter technique was presented by
Letarte et al. (2002), covering the whole galaxy and identifying 904
carbon stars. A similar number of M-stars with colours corresponding
to spectral type M0 and later was identified. Unfortunately no
information is given on the number distribution of M-stars as a
function of colour (spectral type). I took their $R,I$ data, and
calculated the bolometric magnitudes adopting $E(B-V)$ = 0.24 and BCs
from CF.

\underline{Sagittarius DIG}

\noindent
A new survey using the narrow-band filter technique was presented by
Demers \& Battinelli (2002), covering the whole galaxy. They found 16
carbon stars. The foreground contamination is so strong that no useful
estimate of the number of M stars could be made. I took their $R,I$
data and reddening ($E(B-V)$= 0.05) and the BCs from CF to estimate
the bolometric magnitudes.

\underline{Pegasus}

\noindent
A new survey using the narrow-band filter technique was presented by
Battinelli \& Demers (2000), covering the whole galaxy. They found 40
carbon stars for their preferred reddening of $E(B-V)$ = 0.03.  The
C/M ratio they quote appears to be in error however.  Based on their
Figure~4, I count 19 M-stars in the area of approximately 5.6
sq.arcmin they define as the foreground area, and 116 M-stars in the
11.4 sq.arcmin covered by Pegasus. This implies there are 11.4/5.6
$\times$ 19 = 39 foreground stars in the area covered by Pegasus, and
hence the C/M ratio is 40 /(116 - 39) = 0.52, and not 0.77. Assuming
that 57\% of all M-stars in the field are foreground, I estimated the
number of M0+, M2+, M3+, M5+, M6+ stars in Pegasus to be 77, 30, 17,
1, 0, respectively. I calculated the bolometric magnitude using the
BCs from CF.

\underline{Aquarius}

\noindent
A new survey using the narrow-band filter technique was presented by
Battinelli \& Demers (2000), covering the whole galaxy. They found 3
carbon stars. The foreground contamination is such that no reliable
estimate for the C/M ratio can be made. I calculated the bolometric
magnitude using the BCs from CF, and a reddening of $E(B-V)$ = 0.03.

\underline{Tucana}

\noindent
A new survey using the narrow-band filter technique was presented by
Battinelli \& Demers (2000), covering the whole galaxy. They found no
carbon stars.

\underline{IC 1613}

\noindent
A new survey, covering the whole galaxy, using the narrow-band filter
technique is presented by Albert et al. (2000). They identify 195
carbon stars, and, considering the foreground contamination, about 300
stars of spectral type M0 and later. Assuming the same fraction of
foreground contamination as for all M-stars, I estimated from their
Figure~5, that there are approximately 35 M5+ and 2 M6+ stars. I took
their $R,I$ data, and calculated the bolometric magnitudes assuming
$E(B-V) = 0.03$ and BCs from CF.

Kurtev et al. (2001) identify the first known Mira in IC 1613, with
a period of 640 days, and an amplitude in the $R$-band of 2.5-3
magnitudes. Its spectral type is around M3.

\underline{WLM}

\noindent
The colours of the M-stars in Cook et al. (1986) were inspected and
I estimated the number of M0+, M2+, M3+, M5+ stars to be 6, 3, 3, 0.

\underline{NGC 55}

\noindent
I corrected an error in the distance and $M_{\rm V}$ in G99.  The
colours of the M-stars in Pritchet et al. (1987) were inspected
and the number of M3+, M5+ stars is estimated to be 7 and 6.

\underline{NGC 300}

\noindent
Richer et al. (1985) identify 25 M-stars (with an estimated 2
foregound objects) with observed colours $V-I \approx 1.9$ and assume
all to be of spectral type M5+. Using Table~1, I estimated the number
of stars in NGC 300 with spectral type M1+, M2+, M3+, M5+ to be 23,
21, 15, 6, respectively.

\vspace{5mm}

\noindent
For the Sextans dwarf galaxy, NGC 205 and NGC 2403 there has been no
new information on the carbon star population since G99.

\vspace{5mm}
\noindent
Table~2 summarises the number of known carbon stars in external
galaxies. Local Group members not explicitly mentioned have no
published information on their C-star population. The last three
entries are galaxies outside the Local Group. Listed are the adopted
distance, absolute visual magnitude, metallicity (these three
parameters come from Mateo (1998) and van den Bergh (2000)), number of
known carbon stars, the surface area on the sky of the respective
survey, the ratio of the area of the galaxy on the sky (taken from the
NED catalogue) and the survey area, and the number of late-type
M-stars, when known (the symbol `3+' meaning stars of spectral type M3
and later, etc.). The entry for the number of carbon stars in our
galaxy is the local surface density of TP-AGB stars by Groenewegen et
al. (1992).

\begin{table}[t]
\caption[]{The carbon star luminosity function. Galaxies are ordered
by decreasing $M_{\rm V}$}
%\vspace{-2mm}
%\small

\begin{flushleft}
\begin{tabular}{lrrrrrr} \hline
Name & $M_{\rm bol}^{\rm max}$ & $M_{\rm bol}^{\rm min}$ & 
$M_{\rm bol}^{\rm mean}$ & spread & $N_{\rm all}$ & $N_{\rm M_{bol} > -3.5}$ \\
     & (mag) & (mag) & (mag) & (mag) &  &  \\ \hline
M 31    &  $-$6.13 &  $-$3.34 &  $-$4.31 &   0.50 &  243 &   16 \\
NGC 2403&  $-$6.05 &  $-$5.89 &  $-$5.97 &   0.08 &    4 &    0 \\
NGC 300 &  $-$5.93 &  $-$4.68 &  $-$5.39 &   0.41 &   13 &    0 \\
LMC     &  $-$8.01 &  $-$1.33 &  $-$4.67 &   0.61 & 7650 &  378 \\
NGC 55  &  $-$4.98 &  $-$3.37 &  $-$4.40 &   0.51 &    9 &    1 \\
NGC 205 &  $-$5.14 &  $-$4.04 &  $-$4.50 &   0.44 &    7 &    0 \\
SMC     &  $-$8.17 &  $-$1.62 &  $-$4.33 &   0.81 & 1626 &  226 \\
NGC 6822&  $-$6.48 &  $-$3.65 &  $-$5.00 &   0.44 &  904 &    0 \\
IC 1613 &  $-$6.34 &  $-$3.48 &  $-$4.94 &   0.56 &  195 &    2 \\
SagDSph &  $-$4.37 &  $-$3.11 &  $-$3.81 &   0.40 &   16 &    5 \\
Fornax  &  $-$5.22 &  $-$3.64 &  $-$4.28 &   0.36 &   41 &    0 \\
Pegasus &  $-$5.59 &  $-$4.05 &  $-$4.68 &   0.37 &   40 &    0 \\
SagDIG  &  $-$5.66 &  $-$4.48 &  $-$5.08 &   0.31 &   16 &    0 \\
Leo I   &  $-$5.38 &  $-$3.32 &  $-$4.59 &   0.47 &   23 &    1 \\
And II  &  $-$4.11 &  $-$3.24 &  $-$3.68 &   0.62 &    2 &    1 \\
Aquarius&  $-$5.12 &  $-$4.71 &  $-$4.89 &   0.21 &    3 &    0 \\
Leo II  &  $-$4.26 &  $-$3.20 &  $-$3.75 &   0.37 &    6 &    1 \\
Sculptor&  $-$4.49 &  $-$2.71 &  $-$3.22 &   0.57 &    8 &    7 \\
Phoenix &  $-$3.71 &  $-$3.55 &  $-$3.63 &   0.11 &    2 &    0 \\
Carina  &  $-$4.85 &  $-$3.14 &  $-$4.30 &   0.61 &    6 &    1 \\
Ursa Minor&  $-$2.90 &  $-$2.04 &  $-$2.60 &   0.33 &    6 &    6 \\
Draco   &  $-$3.62 &  $-$2.92 &  $-$3.27 &   0.35 &    3 &    2 \\
\hline
\end{tabular}
\end{flushleft}
\vspace{-1mm}
\end{table}

\begin{figure}[t]
\centerline{\psfig{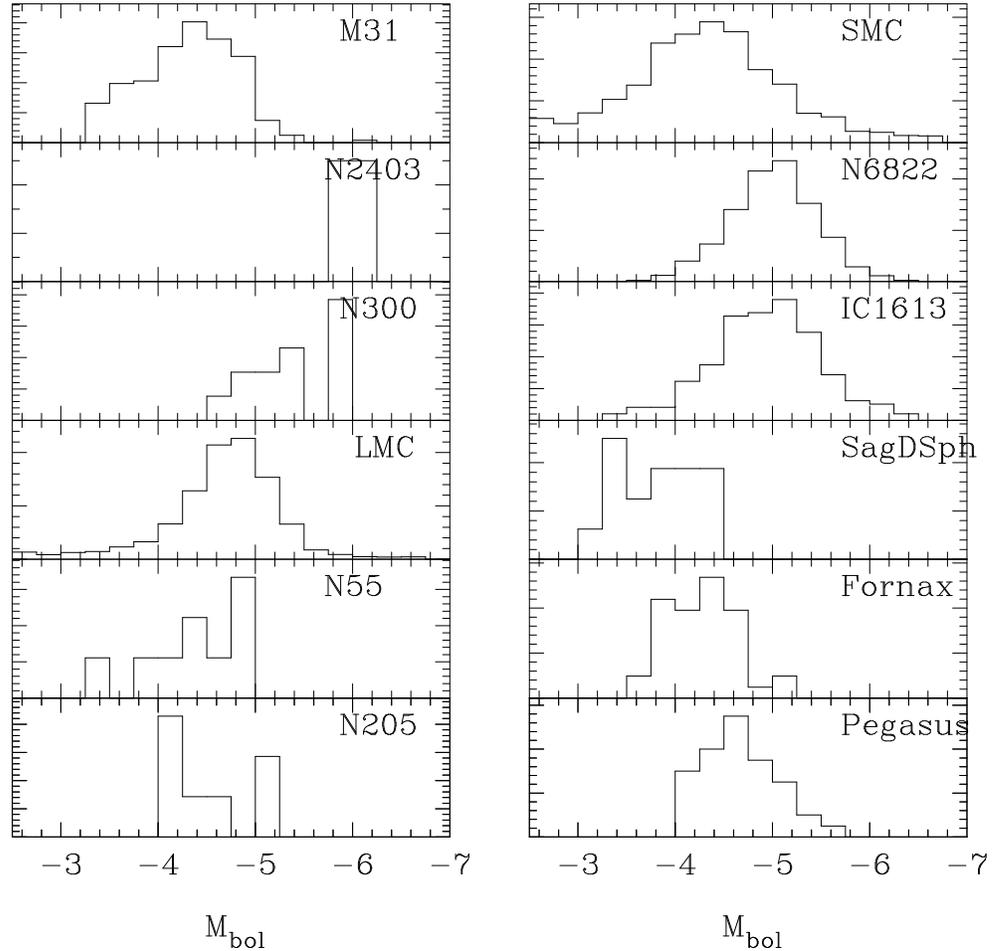}}
\caption[]{Luminosity functions, ordered by decreasing $M_{\rm V}$ of
the galaxies, ordered top to bottom, left to right.  The total number
of stars plotted per galaxy is listed in column 6 of Table~3. In the
case of the SMC and LMC, the lowest luminosity bin is cumulative.}
%\vspace{-3mm}
\end{figure}

\setcounter{figure}{6}
\begin{figure}[t]
\centerline{\psfig{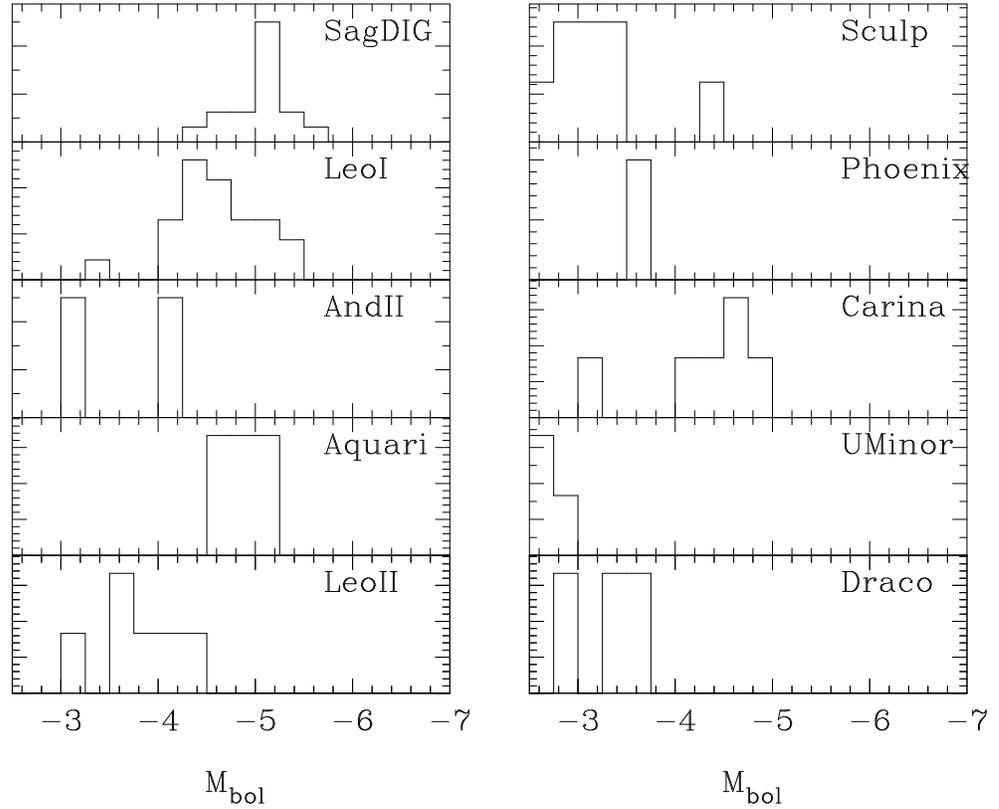}}
\caption[]{Continued.}
%\vspace{-3mm}
\end{figure}

\begin{figure}[t]
\centerline{\psfig{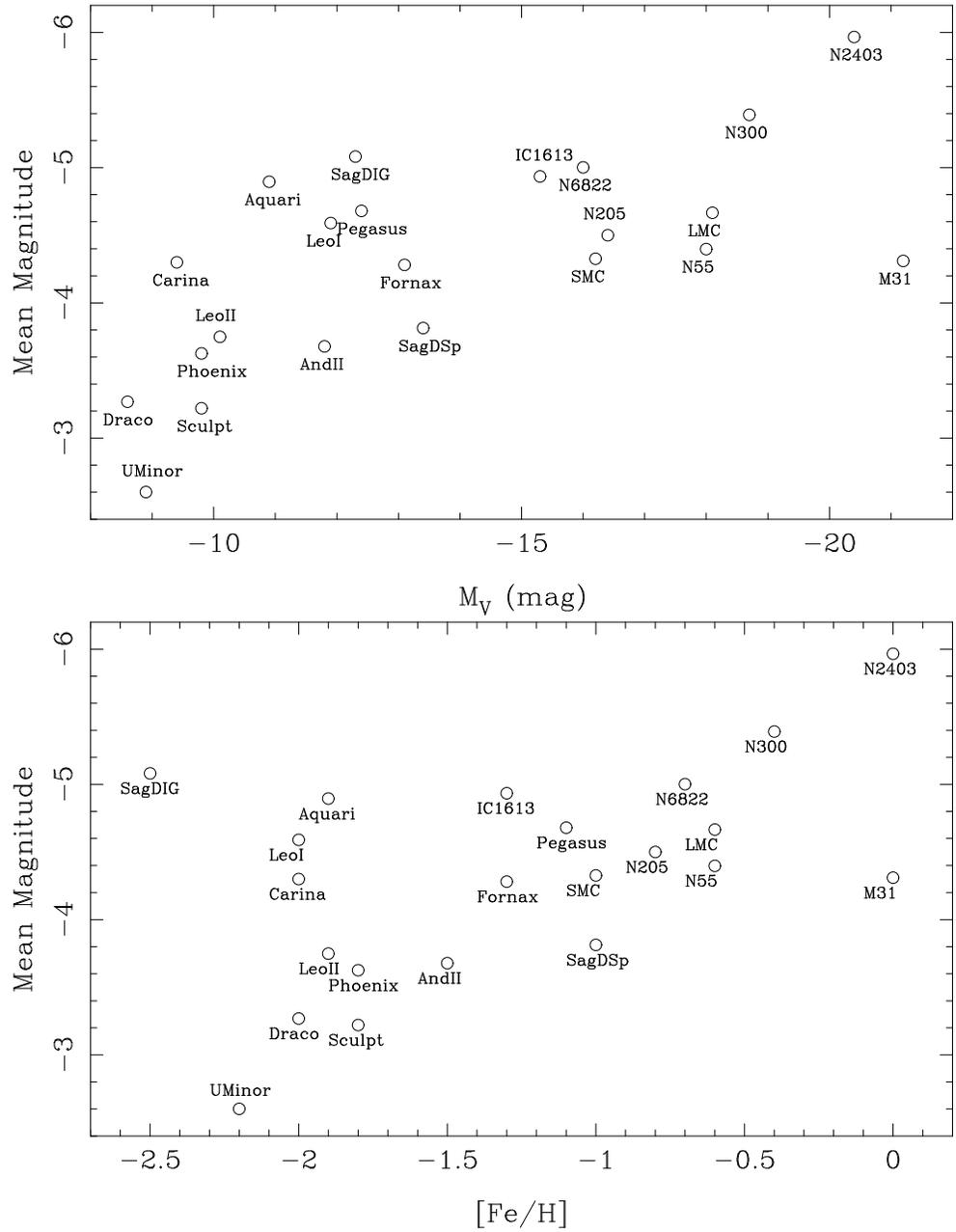}}
\caption[]{Mean bolometric magnitude of carbon stars versus 
$M_{\rm V}$ and metallicity.}
%\vspace{-2mm}
\end{figure}

\section{Discussion}

Figure~3 shows the well-known correlation between metallicity and
absolute visual magnitude. The interpretation being that the more
luminous galaxies are also the more massive ones that have had more
gas mass available to transform into stars and enrich the interstellar
medium. There is a clear outlier, SagDIG, but several research groups
find a similar low metallicity for this galaxy (see discussion in
Momany et al. 2002)

Figure~4 shows the number of carbon stars in a galaxy versus $M_{\rm
V}$ represented in two ways. First, the total number of C-stars in a
galaxy was estimated by simply multiplying the known number by the
ratio of total surface area of a galaxy to the survey area. As for
some galaxies the survey area is less than a few percent, this
correction factor can be quite large (and uncertain).  To circumvent
this, the bottom panel shows the surface density of carbon stars in
the particular survey. The drawback of this approach is that it does
not take into account the spatial variation of the density of carbon
stars within a galaxy. In neither approach did I correct for the fact
that we do not see these galaxies face-on. Some interesting things can
be noticed. There is a clear relation between the (estimated) total
number of C-stars and $M_{\rm V}$, and there seems to be a maximum
surface density of about 200 kpc$^{-2}$ averaged over a galaxy. In
both plots NGC 55, 300 and 2403 are clear outliers. These are the most
distant galaxies surveyed, and one might suppose that the surveys have
been incomplete. For NGC 55 the explanation probably lies as well in
the fact that we see this galaxy almost disk-on, and so both the total
number as well as the surface density have been
underestimated. Reddening within the galactic disk of the galaxy can
also play a role. For NGC 2403 the small number of carbon stars lies
in the fact that the survey has been incomplete. All 4 known C-stars have
luminosities that are much higher than the average in galaxies for
which we know the LF in more detail. The same is true for NGC 300.  A
last note of caution is that I did not try to make a distinction
between carbon stars on the TP-AGB and lower-luminosity carbon
stars. For example, the data for our Galaxy represents TP-AGB stars
only, while on the other hand the SMC is known to contain a large
fraction of low-luminosity carbon stars (see later).

Figure~5 shows the ratio of C- to late M-stars, and it confirms the
well known trend (Cook et al. 1986, Pritchet et al. 1987). The
interpretation is that a star with a lower metallicity needs fewer
thermal pulses to turn from an oxygen-rich star into a carbon star.

Figure~6 shows the ratio of the total estimated number of carbon stars
over the visual luminosity of the galaxy (e.g. Aaronson et al. 1983). Most
of the galaxies scatter between a value of --3 and --4, with a few
outliers which are the same as noticed in Fig.~2.
 
Figure~7 shows the C-stars bolometric LF for the galaxies for which it
could be constructed (Sect.~2); for the Magellanic Clouds see
Groenewegen (1998) for details. Table~3 lists the maximum, minimum and
mean magnitude, as well as the spread, calculated from the rms
deviation from the mean.  Also listed are the number of stars that
went into the calculation and are plotted in Fig.~7, and the number of
C-stars that are fainter than $M_{\rm bol} = -3.5$, the luminosity of
the tip of the RGB.  This is not an absolute foolproof limit between
true AGB stars and binary masqueraders as (low initial mass) genuine AGB
stars may have a luminosity fainter than $M_{\rm bol} = -3.5$ when
experiencing the first few pulses on the TP-AGB and/or when they
happen to be observed during the luminosity-dip of the thermal pulse
cycle (about 10-20\% of the inter pulse time). As an aside it is noted
that in the case the luminosity is not conclusive about the AGB nature
of an object, one could try to observe one or more Technetium lines
(e.g. Abia et al. 2002 and references therein), as the presence of
this unstable isotope indicates recent nucleosynthesis.

The data show that in well populated LFs, the mean $M_{\rm bol}$ is
between --4 and --5. It also shows that the mean in NGC 300 and NGC
2403 is much higher. Unless one would invoke a large uncertainty in
the distance or a burst of recent star formation, the most natural
explanation lies in the incompleteness of the surveys in these distant
galaxies.  Finally, the data shows that in the fainter galaxies the
mean magnitude increases and that a fair number of C-stars are of the
low-luminosity type. This is more clearly seen in Fig.~8 where the
mean magnitude is plotted. A LF which is dominated by faint carbon
stars is indicative of an absent intermediate age population, and a
corresponding star formation history. It is also evident that there is
a considerable spread at a given metallicity or $M_{\rm V}$,
inhibiting the use of the mean carbon star luminosity as a distance
indicator.

\vspace{5mm}
\noindent
In a recent paper Mouhcine \& Lan\c{c}on (2002) present evolutionary
population synthesis models, including chemical evolution, with
special focus on intermediate age populations. Their models are the
first that are able to account for the observed trend in Fig.~3
adopting `typical' SFH for Sa, Sb, Sc and Irr Hubble type galaxies.
The AGB phase is included through a semi-analytical treatment of the
third dredge-up, with efficiency parameters set to values that have
been determined in other studies to fit the LMC carbon star LF and C/M ratio.

\section{Conclusion}

In principle, the overall carbon star LF and C/M ratio contains
information about the star-formation rate history from, say, 10 Gyr
ago (the low-luminosity C-stars in binaries) to a few-hundred Myr ago
(the high luminosity tail of the LF). Its a challenge to theoretical
models to use these constraints together with other data to derive the
chemical evolution and star formation history of these galaxies. The
models of Mouhcine \& Lan\c{c}on represent a first successful step in
this direction. \\

\acknowledgments

I thank Pierre Royer (KUL) for a critical reading of the manuscript.
This publication makes use of data products from the Two Micron All
Sky Survey, which is a joint project of the University of
Massachusetts and the Infrared Processing and Analysis
Center/California Institute of Technology, funded by the National
Aeronautics and Space Administration and the National Science Foundation.

\end{document}